\documentclass[fleqn,usenatbib]{mnras}

\usepackage{graphicx}	
\usepackage{amsmath}	
\usepackage{amssymb}	
\usepackage{comment}

\graphicspath{{.}}

\title[Supermassive Star Formation via Super Competitive Accretion]
{Supermassive Star Formation via Super Competitive Accretion in Slightly Metal-enriched Clouds}

\author[S. Chon \& K. Omukai]{
Sunmyon Chon $^{1}$\thanks{E-mail: sunmyon.chon@astr.tohoku.ac.jp},
and Kazuyuki Omukai $^{1}$
\\
$^{1}$Astronomical Institute, Graduate School of Science, Tohoku University, Aoba, Sendai 980-8578, Japan
}

\date{Accepted XXX. Received YYY; in original form ZZZ}

\pubyear{2019}

\begin{document}

\label{firstpage}
\pagerange{\pageref{firstpage}--\pageref{lastpage}}
\maketitle

\graphicspath{{../pictures/}}

\begin{abstract}
Direct collapse black hole (DCBH) formation with mass $\gtrsim 10^{5}~M_{\odot}$ 
is a promising scenario for the origin of high-redshift supermassive black holes.
It has usually been supposed that the DCBH can only form in the primordial gas 
since the metal enrichment enhances the cooling ability and causes the fragmentation into smaller pieces.
What actually happens in such an environment, however, has not been explored in detail. 
Here, we study the impact of the metal enrichment on the clouds, 
conducting hydrodynamical simulations to
follow the cloud evolution in cases with different degree of the metal enrichment $Z/Z_{\odot}=10^{-6}-10^{-3}$.
Below $Z/Z_{\odot}=10^{-6}$, metallicity has no effect and
supermassive stars form along with a small number of low-mass stars.  
With more metallicity $Z/Z_{\odot} \ga 5 \times 10^{-6}$, although the dust cooling indeed promotes 
fragmentation of the cloud core and produces about a few thousand low-mass stars,
the accreting flow preferentially feeds the gas to the central massive stars, 
which grows supermassive as in the primordial case.
We term this formation mode as the {\it super competitive accretion},
where only the central few stars grow supermassive while a large number of other stars 
are competing for the gas reservoir. 
Once the metallicity exceeds $10^{-3}~Z_{\odot}$ and metal-line cooling becomes operative, 
the central star cannot grow supermassive due to lowered accretion rate.
Supermassive star formation by the super competitive accretion opens up a new window for seed BHs, 
which relaxes the condition on metallicity and enhances the seed BH abundance.
\end{abstract}


\section{Introduction}
Recent high-redshift quasar surveys uncovered a large population of the supermassive black holes (SMBHs) 
at $z \gtrsim 6$, i.e.,  
when the age of universe was less than $800$ million years
\citep[e.g.][]{Mortlock+2011,Wu+2015,Matsuoka+2016, Venemans+2016, Banados+2018, Onoue+2019}.
Such short timescale poses a challenge on their formation scenarios.
A natural candidate for seed black holes (BHs) are the first-star remnants, 
which were considered to be very massive ranging from a few 10 to a few 100 $M_{\odot}$ 
\citep{Alvarez+2009,Hosokawa+2012, Hosokawa+2016,Johnson+2014, Hirano+2014, Hirano+2015, Susa+2014,Stacy+2016}.
Even for the object at the top end of the mass spectrum $\sim 10^{3}~M_{\odot}$ 
the Eddington-limited accretion must be maintained to reach observed SMBH masses of $10^9~M_{\odot}$ by $z \simeq 6$. 
Duty cycle close to unity for over several orders of magnitude in mass seems improbable 
on the basis that radiation feedback from the accreting BHs easily quenches the efficient mass accretion 
\citep{Milos+2009,PR2011, Sugimura+2018}.
More massive seeds are desirable as the origin of SMBHs.

Formation of massive seeds is accomplished via the so-called direct collapse (DC) \citep[e.g.][for review]{Volonteri2010, Haiman2013, Inayoshi+2019}, 
where a supermassive star (SMS) forms first and then collapses by post-Newtonian instability to a BH with 
mass of $10^{5}-10^{6}~M_{\odot}$
\citep{Shibata+2002, Umeda+2016, Uchida+2017}.
Suppression of H$_{2}$ cooling in the primordial gas is the key for this mechanism to work \citep[e.g.][]{BL2003}.
In this case, clouds in massive enough halos collapse
isothermally at $\sim 8000~$K via atomic cooling \citep{Omukai2001} 
and almost monolithically without major episode of fragmentation \citep{Inayoshi+2014, Becerra+2015}. 
A protostar forms at the center and grows rapidly by accreting the surrounding gas 
at a rate $0.1$--$1~M_{\odot}~\mathrm{yr^{-1}}$.
Such vigorous accretion makes the star inflate in radius as if it were a red giant star, 
which hardly emits ionizing radiation \citep{OmukaiPalla2003, Hosokawa+2012,Hosokawa+2013,Schleicher+2013,Haemmerle+2018}.
Radiation feedback onto the surrounding flow being negligible, 
the accretion continues until the star becomes supermassive and directly collapses to a BH 
\citep[e.g.][]{Chon+2018}.

Among candidate sites for the DC, most widely studied are metal-free atomic-cooling halos irradiated 
with intense far-ultraviolet (FUV) radiation
from neighboring halos \citep[e.g.][]{Dijkstra+2008, Chon+2016, Chon+2017, Regan+2017, Maio+2019}. 
Required level of FUV radiation, however, turned out to be too high 
\citep[e.g.][]{Shang+2010,WolcottGreen+2011,Sugimura+2014} to account for all the seeds 
of SMBHs ubiquitously residing in galaxies in the present-day universe although 
this mechanism may be able to explain the existence of very high-$z$ rare SMBHs.  
Other environments for DC have also been proposed including
dense clouds experiencing shock heating \citep{Inayoshi+2012}, 
halos with large streaming motions between baryon and dark matter \citep{Hirano+2017b}, 
halos subject to strong dynamical heating due to frequent mergers \citep{Wise+2019}, etc.. 
Those mechanisms, however, more or less share common short comings with the FUV scenario:
the expected number of seeds can account for only a small number of rare high-$z$ SMBHs, 
rather than being universal origin for all the SMBHs. 

It has been believed that the DC takes place only in a primordial gas 
since in a gas with even slight metal enrichment vigorous fragmentation due to dust cooling 
leads to formation of a star cluster, rather than an SMS \citep{Omukai+2008, Latif+2016a}.
Although a very massive star could form through runaway collision of stars in a dense star cluster, 
its mass is predicted to be at most $10^{3}~M_{\odot}$ \citep{Katz+2015,Sakurai+2017,Reinoso+2018}, 
not well enough as a seed for high-$z$ SMBHs. 
One important aspect, the presence of a massive gas inflow, however, has been missed 
in the those $N$-body stellar dynamics calculations.  

In this paper, we investigate star cluster formation in clouds both with slight metal enrichment 
and with strong FUV irradiation to find out how massive the stars become both by gas accretion and merger. 
By way of high-resolution three dimensional simulation resolving down to the $\sim$au scale and 
introducing sink particles for stars, 
we follow the entire evolution of the collapse and fragmentation of the cloud, 
and long-term evolution of the stellar system and mergers among the members.
We indeed observed that vigorous fragmentation occurs once the dust cooling 
becomes effective at $Z/Z_{\odot} \gtrsim 5 \times 10^{-6}$. 
Nevertheless, the central star grows to supermassive in a runaway fashion via combination of gas accretion and stellar mergers. 
The evolution of the central star is almost the same as what is envisaged in the DC scenario despite the metal enrichment.
Above $Z/Z_{\odot} \sim 10^{-3}$, the central stars cannot grow supermassive any more
as the metal-line cooling prohibits the formation of SMS by lowering the accretion rate.
Recently, \citet{Tagawa+2019} also claimed from analytical argument that 
the central star forming in a dense star cluster grows via the gas accretion as well as 
the stellar collision and becomes an SMS of $10^{4}$--$10^{5}~M_{\odot}$. 
The result of our numerical calculation agrees with their expectation. 

This paper is organized as follows. We describe the numerical methods in Section~\ref{sec::methodology}.
In Section~\ref{sec::results}, we present our main results.
In Section~\ref{sec::discussion}, we discuss uncertainties and implications of our results. 
Finally, summary and concluding remarks are given in Section~\ref{sec::summary}.

\begin{figure}
	\centering
		\includegraphics[width=8.cm]{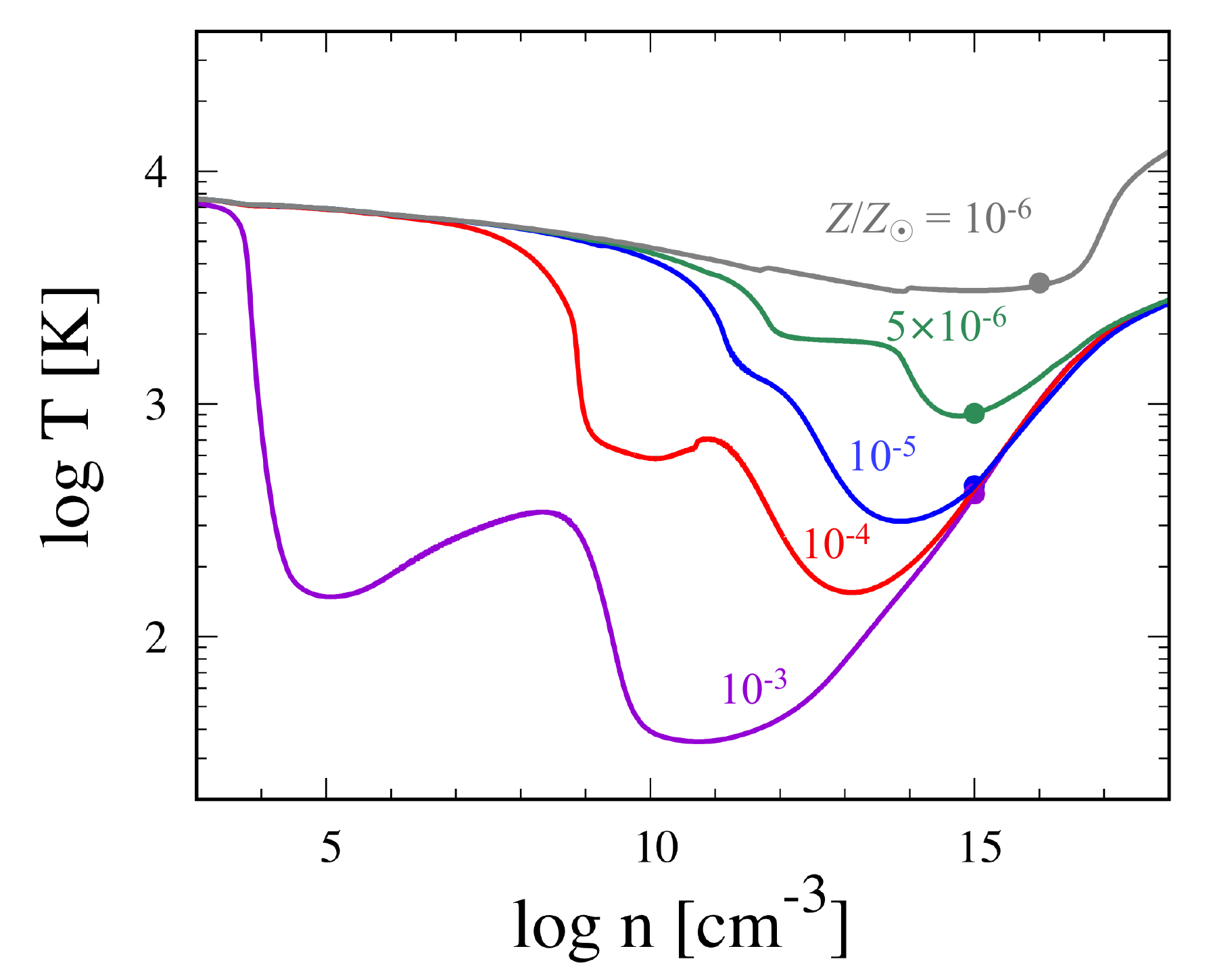}
		\caption{The temperature as a function of density, 
                or the equation of state, used in our calculation.
		We adopt the results obtained by the one-zone calculations
		for the five different metallicities with $Z/Z_{\odot} = 10^{-3}$ (purple), 
		$10^{-4}$ (red), $10^{-5}$ (blue), $5\times 10^{-6}$ (green), and $10^{-6}$ (grey)
		under the strong FUV radiation \citep{Omukai+2008}. 
		The bold circles indicate the threshold density $n_\text{ad}$,
		above which we switch to the stiff (``adiabatic'') equation of state, $p \propto \rho^{5/3}$.
		}
		\label{fig_eos}
\end{figure}

\begin{figure*}
	\centering
		\includegraphics[width=17.5cm]{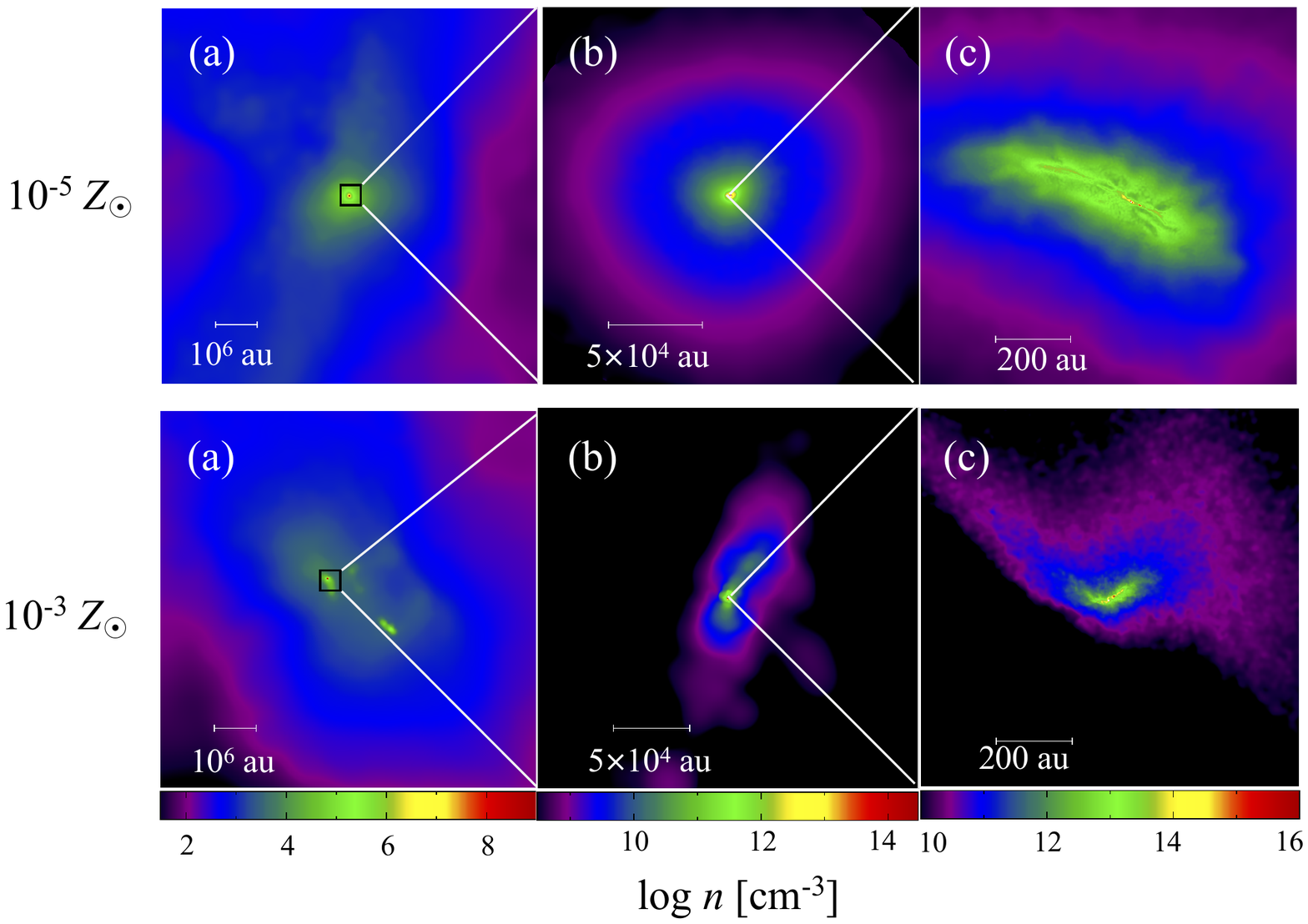}
		\caption{Density distribution around the cloud center just before the formation 
                of the primary sink particle for metallicities $Z/Z_{\odot}=10^{-5}$ (upper) and $10^{-3}$ (bottom panels).
		}
		\label{fig_snapshot_coll}
\end{figure*}

\begin{figure*}
	\centering
		\includegraphics[width=17.5cm]{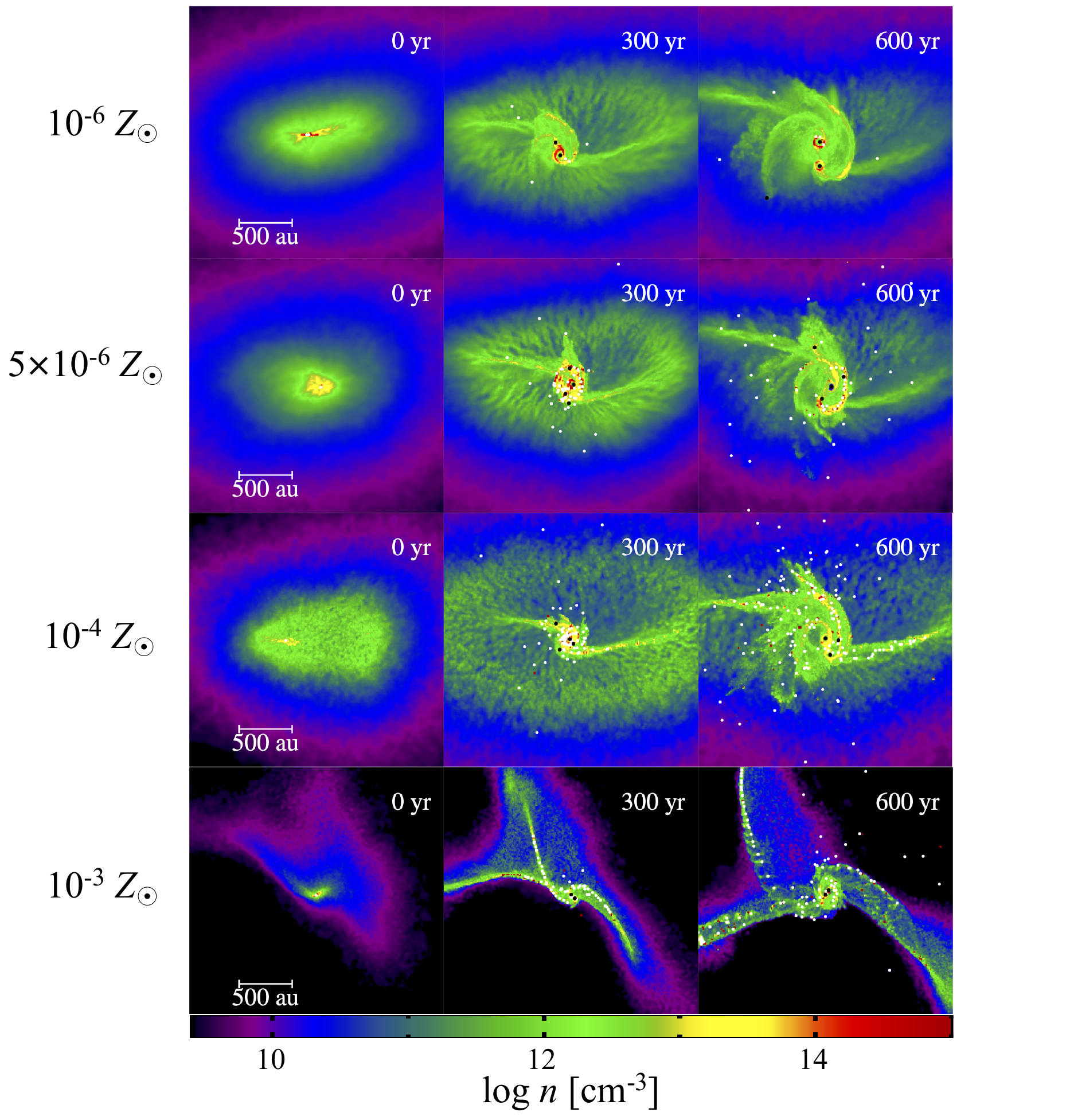}
		\caption{Density distribution around the primary sink particle 
		for metallicities $Z/Z_{\odot} = 10^{-6}$, $5\times10^{-6}$, $10^{-4}$, and $10^{-3}$ 
                (from top to bottom).
		The result of the case with $Z/Z_{\odot} = 10^{-5}$ is very similar 
                to that of $10^{-4}$ and we do not present it here.
		Rows represent the time sequences at $t=0$, i.e., the epoch of the primary sink formation, 
                $300$, and $600~$years after that. 
		Black circles represent massive sink particles with $M_{*} > 10~M_{\odot}$, 
		while white ones those with lower mass $M_{*} < 10~M_{\odot}$.}
		\label{fig_snapshot}
\end{figure*}

\section{Methodology} 
\label{sec::methodology}
We perform a suite of hydrodynamics simulations using the smoothed particle hydrodynamic (SPH) code, 
{\tt Gadget-2} \citep{Springel2005}.
We extract the volume around the halo labeled ``Spherical Cloud'' from the cosmological simulation of \citet{Chon+2018} 
as the initial condition of our calculation. 
This halo is suitable for this study, 
in a sense that it is a typical DC site, which is located close to a luminous galaxy and 
irradiated by an intense UV radiation. 
\citet{Chon+2018} in fact have found SMS formation occurring via the DC in this halo in the case of metal-free composition.
Here, we will investigate the impact of the metal pollution on the cloud evolution.

Our initial simulation volume contains the gas particles inside 
$7 \times 10^{6}~$au from the cloud center.
After the central protostar is formed, 
we extract again the particles inside $10^{5}~$au, 
containing the total gas mass of $\sim 2 \times 10^{4}~M_{\odot}$,
and resume the calculation.
We conduct particle splitting following \citet{Kitsionas+2002},
to resolve the Jeans length with sufficient resolution \citep{Truelove1997}.
The splittings are performed at the four refinement densities of  
$10^{6}$, $10^{8}$, $10^{10}$, and $10^{14}~\mathrm{cm^{-3}}$.
The initial particle mass is $1.6~M_{\odot}$,
while the particle mass in the highest splitting level is $5.5 \times 10^{-5}~M_{\odot}$.
This assures us to safely resolve the forming protostellar cores.

The equation of state (EoS) of the gas is assumed to follow the barotropic relation, i.e., 
the pressure is given by a function of the gas density,
which is pre-calculated by the one-zone model \citep{Omukai+2008}, 
where the clouds with different metallicities are irradiated by sufficiently strong FUV radiation
for DC.
The temperature remains about $8000~$K by the atomic cooling as the molecular hydrogen is dissociated
until metal-line or dust cooling effect takes place. 
Fig.~\ref{fig_eos} shows the adopted temperature evolution as a function of density for five metallicities studied, 
$Z/Z_{\odot} = 10^{-3}$ (purple), $10^{-4}$ (red), $ 10^{-5}$ (blue), $5\times10^{-6}$ (green), and $10^{-6}$ (grey).
Note that the EoS with $Z/Z_{\odot} = 10^{-6}$ is identical to that of the primordial one. 
The case with $5\times10^{-6}~Z_{\odot}$ corresponds to the threshold metallicity, above which 
the dust cooling makes the temperature suddenly drop at high density $\gtrsim 10^{10}~\mathrm{cm}^{-3}$. 
With metallicity as high as $Z/Z_{\odot} = 10^{-3}$, the temperature drops at a low density $\sim 10^4~{\rm cm^{-3}}$
due to the metal-line cooling.
Above a high enough density $\rho_\text{ad}$, shown by solid circles denoted in Fig.~\ref{fig_eos}, we set 
the EoS stiff or ``adiabatic'' as $p \propto \rho^{5/3}$, where $p$ and $\rho$ are the gas pressure and the density, 
respectively, to avoid very short time step at very dense collapsed regions. 
The numerical values for $n_\text{ad} \equiv \rho_\text{ad}/ m_\text{p}$ are $10^{15}~\mathrm{cm}^{-3}$ 
for the metallicities with $Z/Z_{\odot} = 10^{-3}$, $10^{-4}$, $ 10^{-5}$, and $5\times10^{-6}$
and $n_\text{ad} = 10^{16}~\mathrm{cm}^{-3}$ for $Z/Z_{\odot} = 10^{-6}$, where $m_\text{p}$ is the proton mass.
Radiative feedback from the stars is not important in our calculation except in few cases
due to high accretion rate and/or small mass of stars and not taken into account (see discussion). 

Once the density exceeds $\rho_\text{sink}$, we introduce a sink particle assuming the protostar is formed there.
The sink particles are created at the density $\rho_\text{sink} = 20 \rho_\text{ad}$, 
with its initial radius of $1~$au, 
which assures the sinks to be placed only at self-gravitating cores.
Protostars accreting at rates higher than $0.04~M_{\odot}{\rm yr}^{-1}$ inflate in radii with 
\begin{align} \label{eq::stellar_radius}
R_{*} = 12~\mathrm{au} \left ( \frac{M_{*}}{100~M_{\odot}} \right ) ^{1/2},
\end{align}
where $M_{*}$ is the mass of the star \citep{Hosokawa+2013}.
We set the sink radius to be $R_{*}$ if the accretion rate is higher than $0.04~M_{\odot}~\mathrm{yr}^{-1}$.
We allow the sink particles to merge each other once the separation between them
becomes smaller than the sum of the sink radii.

\section{Results} 
\label{sec::results}
\subsection{Early evolution: cloud collapse until the protostar formation}
The way that the cloud collapse proceeds depends largely on whether the metal-line or dust cooling becomes 
important at some density. 
Fig.~\ref{fig_snapshot_coll} shows morphology of the clouds with $Z/Z_{\odot} = 10^{-5}$ (top) and $10^{-3}$ (bottom)
when the maximum density reaches $n=10^{16}~\mathrm{cm^{-3}}$ and the first sink is forming. 
The former is an example of cases with only dust cooling, while the latter is that with both 
the dust and metal-line cooling. 
The first sink formed in each run, which will also grow to the most massive one, is referred to as the ``primary star'', 
hereafter.

First, let us see the case of $Z/Z_{\odot} = 10^{-5}$. 
The overall morphology of the collapsing clouds is very similar among the cases with 
metallicity $Z/Z_{\odot} \lesssim 10^{-4}$, 
where the metal-line cooling is absent although 
dust cooling becomes effective at high densities $\sim 10^{10}~\mathrm{cm^{-3}}$ except the case with 
$Z/Z_{\odot} = 10^{-6}$.
The cloud initially maintains spherical shape. 
Once the dust cooling becomes operative, 
it then starts to be elongated and becomes filamentary in shape (panel c) 
as a result of rapid growth of bar-mode density perturbations in the case of decreasing temperature with increasing density 
\citep[e.g.][]{Tsuribe+2006, Chiaki+2016, Chon+2018}.

With metallicity $Z/Z_{\odot} = 10^{-3}$,
the metal-line cooling becomes effective at much lower density $\sim 10^{4}~\mathrm{cm^{-3}}$.
As a result of this,  
the cloud fragments at much lower density with a scale of $\gtrsim$ pc (panel a).
The density distribution at smaller scales $<10^{4}~$au is also dramatically changed:
the size of the central dense core becomes much smaller and the density in the surroundings much lower, 
which will result in lower accretion rate onto the primary protostar after its formation 
(see Section~\ref{sec::LongTermEvolution}).

\subsection{Late evolution: star cluster formation} \label{sec::LongTermEvolution}
\begin{figure}
	\centering
		\includegraphics[width=8.2cm]{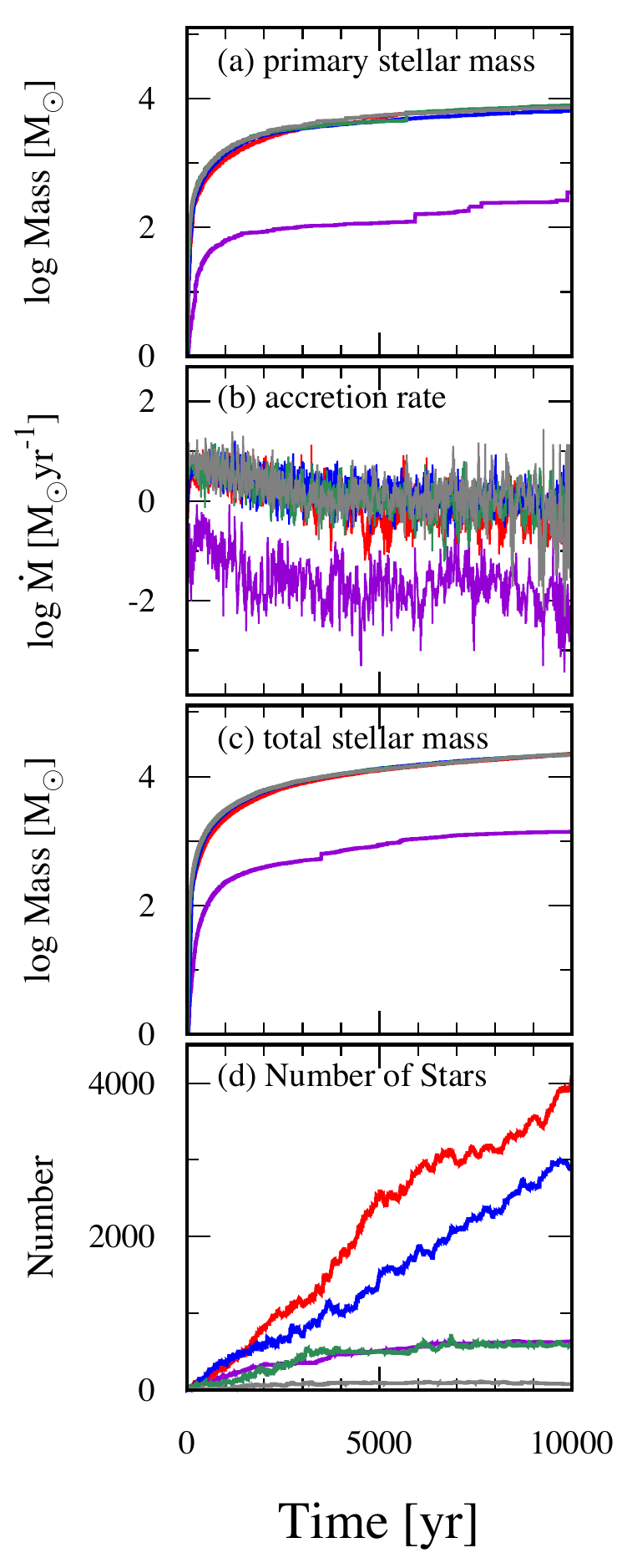}
		\caption{Time evolution of (a) the primary star mass, 
		(b) the mass accretion rate onto the primary star,
		(c) the total mass of stars, 
		and (d) the number of the stars present in the system
		for the cases with $Z/Z_{\odot}=10^{-3}$ (purple), $10^{-4}$ (red), 
                $10^{-5}$ (blue), $5\times10^{-6}$ (green), and $10^{-6}~Z_{\odot}$ (grey).
                The time is measured since the primary sink formation.
		}
		\label{fig_SF_evo}
\end{figure}

We then describe the evolution after the primary star formation until the end of our calculation, i.e., 
$10^{4}~$years after its formation.
Fig.~\ref{fig_snapshot} shows the density distributions around the primary star
at $t=0$, $300$, and $600~$years after its formation 
for metallicities $Z/Z_{\odot}=10^{-6}, 5 \times 10^{-6}, 10^{-4}$, and $10^{-3}$.
The case with $Z/Z_{\odot}=10^{-5}$ is very similar to that with $10^{-4}$ and not presented here.
Fig.~\ref{fig_SF_evo} shows the evolution of (a) the mass of the primary star, 
(b) the mass accretion rate onto the primary star, (c) the total mass of stars, and 
(d) the number of stars still surviving, as functions of time after the primary star formation
for all the five metallicities studied. 

~

Overall evolutionary features can be summarized as follows:

(i) With metallicity as low as $10^{-6}~Z_{\odot}$ (top row), 
the temperature evolution being identical as in the primordial case, 
only a small number of fragments are formed in the circumstellar disk, as in the ordinary DC cases
\citep[e.g.][]{Latif+2013, Inayoshi+2014, Sakurai+2016, Shlosman+2016, Chon+2018, Luo+2018, Matsukoba+2019}.
The primary star grows to the mass of $\sim 10^{4}~M_{\odot}$ in $10^{4}~$years.

(ii) With slight metal-enrichment (second and third rows), 
the dust cooling becomes important at high density and induces fragmentation of the circumstellar disks
\citep[e.g.][]{Clark+2008,Dopcke+2013,Tanaka&Omukai2014}. 
The density structure is modified only at scales smaller than $100~$au
and it remains very similar to that in the DC case at larger scales.
The mass of the primary star reaches $\sim 10^{4}~M_{\odot}$ also in this calculation.  
The number of stars increases with increasing metallicity from 600 at $5 \times 10^{-6}~Z_{\odot}$ 
to 4000 at $10^{-4}~Z_{\odot}$.

(iii) When the metallicity becomes as high as $10^{-3}~Z_{\odot}$ (bottom row),
the metal-line cooling becomes effective at density $\gtrsim 10^{4}~\mathrm{cm^{-3}}$. 
This results in the reduction of the accretion rate by about two orders of magnitude 
and the primary stellar mass is also reduced to $350~M_{\odot}$.

~

Next, we describe the evolution in each case in more detail:

In case (i) with metallicity $10^{-6}~Z_{\odot}$,
the primary star efficiently grows in mass at a rate of $1$--$10~M_{\odot}~\mathrm{yr}^{-1}$
and attains the mass of $8000~M_\odot$ at $t=10^4~$years.
The circumstellar disk forms around the primary star at $t=300~$years,
and then fragments by the gravitational instability, 
forming multiple stars. Around 80 stars have been formed by $t=10^4~$years.
Eventually a few fragments grow massive and form a stable binary system with the primary star.
This picture is consistent with the previous calculation by \citet{Chon+2018},
who start calculation from the same initial condition and find that an SMS binary system is 
formed in this sample.
We note, however, that since our simulation here has higher spatial resolution
and is able to capture finer structures around the protostars,
the number of the stars formed is also larger than \citet{Chon+2018}'s result
\citep[e.g.][]{Machida&Doi2013,Susa2019}.

In case (ii) with metallicity $5 \times 10^{-6}~Z_{\odot}-10^{-4}~Z_{\odot}$, 
the behaviors shown in panels (a)-(c) are very similar to each other 
and also to the DC case (i) of $10^{-6}~Z_{\odot}$.
The mass of the primary star reaches $6 - 8\times 10^{3}~M_{\odot}$ by $t = 10^{4}~$years with
accretion rate $1$--$10~M_{\odot}\mathrm{yr}^{-1}$,
and the total stellar mass is a few times larger than the primary star mass. 
This reflects the fact that the large-scale density structure of $10^{4}$ -- $10^{5}~$au in those cases 
is similar to the primordial DC case as the dust cooling only affects 
the density structures in smaller-scale dense regions. 
In a few hundred years after the primary star formation,
a filamentary structure begins to develop in a compact region of $\sim100~$au
due to the dust cooling (middle column of Fig.~\ref{fig_snapshot} at $300~$years).
The gas is supplied to the primary star through this filament. 
Although fragmentation of the filament produces a number of stars, 
most of them move along the filamentary flow toward the center 
and then merge with the central star.
Since the dynamical timescale at the fragmentation scale is very short ($\lesssim 10^{3}~$years),
the conversion efficiency from the gas to the star is mainly determined by the large-scale gas flow 
with the longer timescale \citep{Li+2003}.
As a result, the central primary star efficiently grows in mass 
despite vigorous fragmentation in the accreting flows due to the dust cooling. 

The filamentary flow, which is twisted due to the angular momentum of the cloud, 
brings the angular momentum to the central region and
the circumstellar disk emerges around the primary star ($t=600~$years).
As the mass accumulates, the disk fragments by the gravitational instability. 
The number of fragments depends strongly on the metallicity: 
only a few fragments are formed in the case with $Z/Z_{\odot}=10^{-6}$,
while almost one hundred fragments appear in the case with $10^{-4}$
by $t=600~$years (Fig.~\ref{fig_SF_evo} d) as a result of more efficient dust cooling.
The fate of fragments formed in the disk is manifold: 
some migrate inward and quickly merge with the central star while others survive and grow somewhat 
by accretion of the gas avoiding merger.
Most of those survivors are ejected from the disk by the tidal interaction with the primary star
while they are still low mass, and they cannot grow anymore after that. 
Therefore, their formation and presence will not be an obstacle for the growth of the central stars, 
which continue to efficiently acquire the accreting gas.

\begin{figure}
	\centering
		\includegraphics[width=8.5cm]{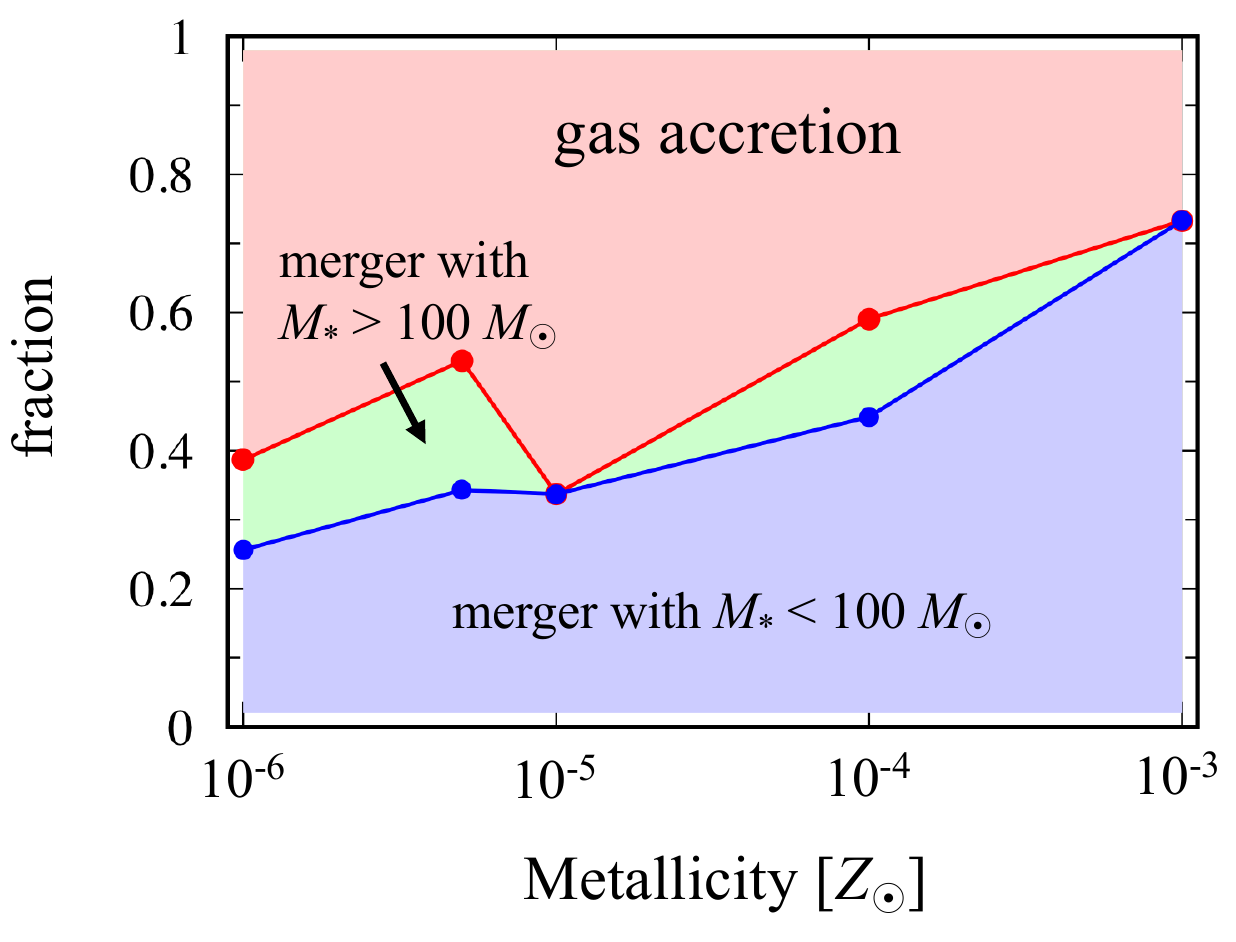}
		\caption{How the primary star acquired mass in cases with different metallicities. 
                The fractions of mass that the primary star obtained
	        by stellar merger with the small stars ($M_{*} < 100~M_{\odot}$; blue) and
		with the massive stars ($M_{*} > 100~M_{\odot}$; green),
		and by gas accretion (red) are shown.}
		\label{fig_mass_origin}
\end{figure}

\begin{figure}
	\centering
		\includegraphics[width=8.cm]{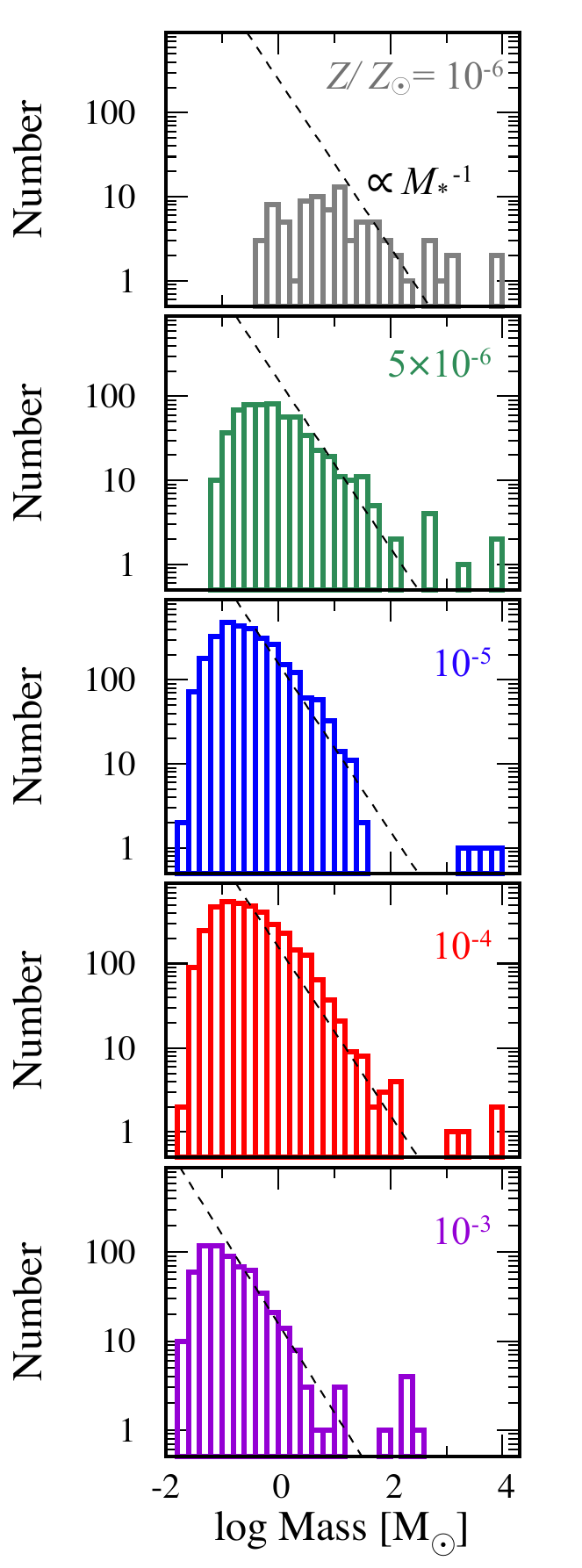}
		\caption{The mass distribution of the stars at $t=10^{4}~$years for the metallicity 
		$Z/Z_{\odot} = 10^{-3}$, $10^{-4}$, $10^{-5}$, $5\times 10^{-6}$, and $10^{-6}$
		from top to bottom.
		Horizontal axis shows the number of stars in each bin, which has logarithmically equal width
                of $\Delta {\rm log} M_{\ast}=0.2$.
		Dashed lines indicate the scaling with $\propto M_{*}^{-1}$.}
		\label{fig_mass_spectrum}
\end{figure}

With metallicity as high as $10^{-3}~Z_{\odot}$ (case iii),
both the mass of the formed stars and the accretion rate onto them become drastically reduced (Fig.~\ref{fig_SF_evo}  a-c).
The primary-star mass reaches only $350~M_{\odot}$ by $t=10^{4}~$years, 
far below those in the lower metallicity cases of $10^{4}~M_{\odot}$.
The total stellar mass is also an order of magnitude smaller.
This is due to the small mass growth rate:
it is initially $\sim 0.1~M_{\odot}~\mathrm{yr}^{-1}$,
but soon declines to $\sim 0.01~M_{\odot}~\mathrm{yr}^{-1}$ at $t \gtrsim 2000~$years.
At this stage, the mass growth proceeds mainly by the stellar merger.
This small accretion rate comes from the lower temperature at $n\gtrsim 10^{4}~\mathrm{cm^{-3}}$ by 
the metal-line cooling since the accretion rate is related to the temperature $T$ as $\propto T^{3/2}$ 
\citep{Shu1977}. 
In fact, the accretion flow at $\sim~$pc scale is largely affected, resulting in 
the reduced inflow rate into the central region.
Although the dust cooling induces fragmentation at $n \gtrsim 10^{10}~\mathrm{cm^{-3}}$, 
the number of stars remains at most several hundred, much smaller than in the cases with $Z/Z_{\odot}=10^{-5}$ and $10^{-4}$
(Fig.~\ref{fig_SF_evo}d).
The final outcome would resemble what has been envisaged as a high-$z$ dense star cluster in the literatures 
\citep[e.g.][]{Omukai+2008,Katz+2015,Sakurai+2017,Reinoso+2018}.

Fig.~\ref{fig_mass_origin} represents the fractions of mass that the primary star acquires through the following three ways: 
gas accretion (red), merger with massive stars ($M_{*}/M_{\odot} > 100$: green),
and merger with small stars ($M_{*}/M_{\odot} < 100$: blue).
We can see that the fraction of mass acquired by merger (accretion) increases (decreases, respectively) with metallicity. 
Note that merger with massive stars tends to be caused by coalescence of 
tight binaries with the initial separation of several $10~$au when one or both of member star(s) 
accrete at high rate and inflate in radii. 
This process is rather stochastic and has no clear trend with metallicity. 

The mass spectra of the stars surviving at the end of our calculation ($t=10^{4}~$years) 
is presented in Fig.~\ref{fig_mass_spectrum}.

First let us see the top end of the mass spectra. 
In the case of $10^{-6}~Z_{\odot}$, supermassive binary stars with $\sim 10^{4}~M_{\odot}$ 
are formed by the DC. 
In slightly metal-enriched cases of $5 \times 10^{-6}, 10^{-5}$, and $10^{-4}~Z_{\odot}$,
we can still see the existence of supermassive stars with $\sim 10^{4}~M_{\odot}$. 
Those components contain the primary star and its binary companions, which 
forms as a result of disk fragmentation. Some of them merge with the primary star.
Once the companion manages to survive for several orbits, however, it can grow more in mass 
than the primary does until their mass ratio becomes close to unity 
and the massive binary system emerges.
This system dominates the gravitational potential of the star clusters and 
thus is preferentially fed by the large-scale accretion flow.
In some cases, although fragmentation occurs in the flow, the small fragments 
do not affect the global flow pattern and most of them are eventually eaten up by the central objects. 
With $10^{-3}~Z_{\odot}$, the most massive objects reach at most a few $100~M_{\odot}$ due to 
the lower accretion rate. They are also a binary with a companion being formed via fragmentation 
of the filamentary flow.

Next, we see mass distribution of the large majority of stars with smaller masses.
In the case of $10^{-6}~Z_{\odot}$, along with the supermassive binary stars, 
about a hundred of less massive stars are formed by disk fragmentation 
with a rather flat mass distribution peaking around $10~M_{\odot}$.
In higher metallicity cases, fragmentation caused by the dust cooling 
leads to the universal mass spectrum of low-mass components 
($M_{*} \lesssim 100~M_{\odot}$), which has 
a peak around $0.1$--$1~M_{\odot}$ and decreases toward higher masses 
with the power-law fashion:
\begin{align} \label{eq::mass_spectrum}
\frac{\mathrm{d} N}{\mathrm{d}\log M_{*}} \propto M_{*}^{-1}.
\end{align}
Note that this power-law exponent agrees with that found by \citet{Bonnell+2001}
for the so-called competitive accretion, where forming protostars compete 
each other for their share of the surrounding gas by the Bondi-like accretion.
Since the mass spectrum at $Z/Z_{\odot}=10^{-3}$ is also similar to those at lower metallicity 
regardless of the effect of metal-line cooling,  
we infer that this universal shape originates from the dust-induced fragmentation.
The fact that the peak mass roughly corresponds to the Jeans mass at $n_\text{ad}$,
which slightly decreases with increasing metallicity, indicates the validity of this interpretation.


\section{Discussion} \label{sec::discussion}
Our results show that the central stellar system efficiently acquires mass and 
SMSs are likely to form when the metallicity is $\lesssim 10^{-4}~Z_{\odot}$.
Here the gas preferentially accretes onto the most massive stars. 
Even when fragments are produced by the dust cooling, 
they move with the inflowing gas and finally merge with the central stars.
For this reason, the primary stellar mass growth is almost independent of the cloud metallicity
as long as the metal-line cooling has negligible effect.
This also explains why the mass growth rate due to the stellar collision is 
much larger than those by 
the runaway collision inside the dense star clusters,
which is driven by the two-body relaxation
\citep[e.g.][]{PZ+2004,Sakurai+2017,Reinoso+2018}.
As our results indicate, the collision time-scale is an order of the free-fall time,
much shorter than that of the two-body relaxation.

Inflation of the stellar radius by accretion also promotes the stellar merger and thus the stellar mass growth.
According to eq.~\eqref{eq::stellar_radius}, the stellar radius becomes several $10$~au at the initial $10^3~$years, 
which exceeds the Jeans length, an order of a few au at the protostar formation.
Since a fragment formed in the disk tends to migrate to an orbit about a Jeans length 
from the central star via the interaction with the gas \citep{Chon+2019},
most of the fragments formed in the disk end up merging with the central star. 
This has already been reported by \citet{Sakurai+2016} in the case of SMS formation in the primordial gas. 

A potential obstacle for stars to continue growing by accretion is the radiative feedback, 
which has been omitted in our calculation.
In our simulations with $Z/Z_\odot \lesssim 10^{-4}$, where the primary stars are expected to grow supermassive, 
they are always accreting the gas at a rate $\gtrsim 0.1~M_{\odot}~\mathrm{yr^{-1}}$. 
Stars with such a high accretion rate inflate in radii and their surface temperature remains as low as several $1000~$K. 
Its UV emissivity would be too small to ionize the surrounding gas 
\citep{Hosokawa+2012,Hosokawa+2013,Sakurai+2015}.
Lower-mass stars with $M_{*}  < 10^{3}~M_{\odot}$ tend to have smaller accretion rate in our simulation
and some of them might experience the Kelvin-Helmholtz contraction and become the main sequence stars.
However, we expect their radiative feedback would have negligibly small impact on the accreting flow
because of their smaller mass and thus the luminosity \citep{Chon+2018}.

We have terminated our calculation at $10^{4}~$years after the primary star formation.
The stars are still growing by accretion at the end of calculation, and 
the stellar growth is expected to continue further.
To determine the final mass of the forming SMSs and their remnant BHs, we need to follow 
the evolution for another million years, 
until the stars collapse either by the post-Newtonian instability 
or by exhausting the nuclear fuel.
In the case of SMS formation by direct collapse in the primordial environment, 
long-term evolution has been followed by several authors 
\citep[e.g.][]{Latif+2013, Sakurai+2016, Shlosman+2016, Chon+2018}.
In their simulations, high-density regions are masked by the sink particles
to save the computational costs.
Those results show that rapid mass accretion indeed continues for $\sim$million years,
as a large amount of gas is trapped around the central massive stars.
We thus expect that our primary stars will follow similar paths of evolution and 
eventually grow to SMSs 
as in the primordial cases \citep[e.g.][]{Latif+2013}.
 
In the case of $Z/Z_\odot = 10^{-3}$, where the metal-line cooling has a significant effect,
the typical accretion rate is smaller than $0.01~M_{\odot}\mathrm{yr}^{-1}$ and 
the stellar feedback would be important before reaching our nominal mass ($350~M_{\sun}$) of the primary star.
As the star contracts to the main sequence at about $100~M_\odot$, 
the ionizing photon emissivity increases rapidly \citep{OmukaiPalla2003}.
The surrounding gas will be photoionized and further accretion is quenched. 
In this case, the likely outcome would be a dense stellar cluster 
with the maximum stellar mass of $\sim 100~M_\odot$. 
This may provide the initial conditions for dynamical evolution of 
dense stellar clusters, where the most massive star is found to 
grow further by stellar merger to several $100~M_\odot$ by way of 
the $N$-body calculations \citep{Katz+2015, Sakurai+2016, Reinoso+2018, Boekholt+2018}.
To determine detailed structure of the cluster,
we need to know how the stellar feedback halts the accretion flow.
Since our motivation in this paper is to examine the possibility of SMS formation,
we leave this issue for a future study.

Previously it has been postulated that the SMS formation by DC 
is only possible in atomic cooling halos both with intense FUV irradiation 
and with primordial gas composition. 
Here we have shown that SMSs can form also in slightly metal-enriched cases 
as long as FUV irradiation is intense enough. 
This relaxation of the condition will increase the expected number density of massive seed BHs. 
Several authors have estimated the number density of such seeds formed in the usual
DC scenario, which requires metal-free gas composition. 
Using the critical intensity advocated by recent studies 
\citep{Sugimura+2014, Agarwal+2014, Inayoshi+2015},
this ranges from a few $\mathrm{Gpc}^{-3}$ \citep[e.g.][]{Dijkstra+2008,Dijkstra+2014}
to $10^{-6}$ -- $10^{-4}~\mathrm{Mpc}^{-3}$ \citep[e.g.][]{Agarwal+2012,Chon+2016,Habouzit+2016,Valiante+2016}.
Although those seeds can account for rare BHs in the high-$z$ universe, 
they fail to be as abundant as all the SMBHs ubiquitously residing in massive galaxies, 
$\sim0.01$--$0.1~\mathrm{Mpc}^{-3}$
\citep{Aller+2002, Davis+2014}.
The DC in the primordial environment is terminated around $z \sim 10$,
as the metal enrichment proceeds \citep[e.g.][]{Trenti+2009, Chon+2016}:
once a Pop~III star ends its life as a core-collapse SN,
metallicity inside the host halo jumps up to $10^{-4}$--$10^{-3}~Z_\odot$
\citep[e.g.][]{Maio+2010,Ritter+2015,Sluder+2016,Chiaki+2018}.
Our calculation demonstrated the SMS formation can continue 
even in a cloud with slight metal-enrichment at later cosmic time.
For example, the first episode of star formation delays another star formation
for a few hundred million years
by the radiative and SNe mechanical feedback, which ejects a gas from the halo.
If the halos approach close enough to a luminous galaxy before another episode of 
star formation, they can be ideal sites for SMS formation. 
Not only the radiation sources outside the halo, but also those inside the same 
halo can trigger SMS formation in an irradiated massive cloud as long as the 
metallicity is low enough. 
With such new varieties of SMS formation sites, the expected seed BH number 
will be largely enhanced. 
We will pursue the validity of this scenario using samples from the cosmological simulations
in the future studies.


\section{Summary}  
\label{sec::summary}
The direct collapse of a cloud is believed to occur, 
leading to formation of a supermassive star in an atomically cooling halo
irradiated by strong FUV radiation in the early universe, 
if the gas is still in the metal-free pristine composition.
We have investigated the impact of metal enrichment on 
star formation in such halos. 
To this end, we have performed hydrodynamical simulations 
for five different metallicities 
$Z/Z_{\odot} = 10^{-6}$, $5\times10^{-6}$, $10^{-5}$, $10^{-4}$, and $10^{-3}$,
by using the temperature evolution pre-calculated by one-zone models. 
Starting from a cosmological halo found in \citet{Chon+2016},
we have followed the evolution for $10^{4}~$years 
after the first protostar is formed at the cloud center.

In almost primordial gas $Z/Z_{\odot} = 10^{-6}$, metal-cooling has no effect 
and the direct collapse ensues, resulting in supermassive star formation with 
accretion rate $\sim 1~M_{\odot}~\mathrm{yr}^{-1}$. 

With slight metal enrichment ($Z/Z_{\odot} = 5\times10^{-6}$, $10^{-5}$, $10^{-4}$), 
clouds fragment by the dust cooling, which makes the temperature drop from several thousand to several hundred K
at high density ($\gtrsim 10^{10}~{\rm cm^{-3}}$), and thousands of stars are indeed formed. 
Majority of them, however, either merge with the central star or are ejected from the system.
As a result, the mass growth history of the central dominant star is not altered from 
the direct collapse case and a supermassive star will be formed along with 
a large number of small stars. 
This is similar to the so-called competitive accretion, which has been observed in some numerical simulations 
of present-day star cluster formation, but much more scaled-up version.  
So we term it as the {\it super-competitive accretion}.

There is a transition in star-cluster formation mode at higher metallicity.
Once the metallicity becomes as high as $Z/Z_{\odot} = 10^{-3}$, 
temperature becomes already as low as 100~K for $\gtrsim 10^{5}~{\rm cm^{-3}}$ by the metal-line cooling. 
Due to the reduced accretion rate $\sim 10^{-2}~M_{\odot}~\mathrm{yr}^{-1}$, the most massive star falls short of 
becoming supermassive, while the smaller stars are continuously formed and grow by accretion 
as envisaged in the ordinary competitive accretion.
\\ \\
This work is financially supported by
the Grants-in-Aid for Basic Research by the Ministry of Education, Science and Culture of Japan 
(SC:19J00324,KO:25287040, 17H01102, 17H02869). 
We conduct numerical simulation on XC50 at the Center for Computational Astrophysics (CfCA) of the National Astronomical Observatory of Japan
and XC40.
We also carry out calculations on XC40 at YITP in Kyoto University.
The work was also conducted using the resource of Fujitsu PRIMERGY CX2550M5/CX2560M5(Oakbridge-CX) 
in the Information Technology Center, The University of Tokyo.
We use the SPH visualization tool SPLASH \citep{SPLASH} in Figs.~\ref{fig_snapshot_coll} and \ref{fig_snapshot}.

\bibliography{biblio2}

\end{document}